\documentclass{aa}
\usepackage{graphicx}
\usepackage{txfonts}
\usepackage{lscape}
\usepackage{longtable, lscape}
\def\HII{H\,{\sc{ii}}}
\def\HI{H\,{\sc{i}}}

\def\HI{H\,{\sc{i}}}
\def\fs{\hbox{$.\!\!^{\rm s}$}}
\def\fdg{\hbox{$.\!\!^\circ$}}
\def\farcm{\hbox{$.\mkern-4mu^\prime$}}
\def\farcs{\hbox{$.\!\!^{\prime\prime}$}}
\def\arcmin{\hbox{$^\prime$}}
\def\arcsec{\hbox{$^{\prime\prime}$}}

\def\degr{\hbox{$^\circ$}}
\def\h{\hbox{$^{\reset@font\r@mn{h}}$}}
\def\m{\hbox{$^{\reset@font\r@mn{m}}$}}
\def\s{\hbox{$^{\reset@font\r@mn{s}}$}}
\def\msol{\hbox{\kern 0.20em $M_\odot$}}

\def\kms{\hbox{\kern 0.20em km\kern 0.20em s$^{-1}$}}
\def\cmmt{\hbox{\kern 0.20em cm$^{-3}$}}
\def\cmmd{\hbox{\kern 0.20em cm$^{-2}$}}
\def\pc{\hbox{\kern 0.20em pc$^{2}$}}

\def\h13cop{\hbox{H$^{13}$CO$^{+}$}}

\begin{document}
   \title{Triggered star formation \\
on the borders of the Galactic H\,{\large{\bf{II}}} region RCW~120 }

%   \subtitle{Evidence for long-distance triggering (ou interactions) ? }

   \author{A.~Zavagno\inst{1}
\and M.~Pomar\`es\inst{1} \and L.~Deharveng\inst{1} \and
    T.~Hosokawa\inst{2}
         \and
         D.~Russeil\inst{1}
    \and
         J.~Caplan\inst{1}\fnmsep\thanks{Based on observations obtained at the European Southern
Observatory using the ESO Swedish Submillimetre Telescope (programme
71.A-0566), on La Silla, Chile}
          }

   \offprints{A. Zavagno}

   \institute{Laboratoire d'Astrophysique de Marseille, 2 place Le Verrier, 13248 Marseille Cedex 4, France
             \and
   Division of Theoretical Astrophysics, National Astronomical Observatory, 2-21-1 Osawa, Mitaka, Tokyo 181-8588,
   Japan \\
              \email{annie.zavagno@oamp.fr}
             }

   \date{Received ; accepted }

% \abstract{}{}{}{}{}
% 5 {} token are mandatory

  \abstract
   {To investigate the process of star formation triggered by the expansion of an \HII\
   region,
   we present a multi-wavelength analysis of the Galactic \HII\ region RCW~120 and its {\mbox{surroundings}}.
   The collect and collapse model predicts that the layer of gas and dust accumulated between the ionization and shock fronts
   during the expansion of the \HII\ region collapses and forms dense fragments, giving rise to potential sites of massive-star formation.}
   {The aim of our study is
   to look for such massive fragments and massive young stars on the borders of RCW~120.}
   {We mapped the RCW~120 region in the cold dust continuum emission at 1.2~mm to search for these fragments. We supplemented this study
   with the available near- (2MASS) and mid-IR (GLIMPSE) data to locate the IR sources observed towards this region and to analyse their properties.
   We then compared the observational results with the predictions of Hosokawa \& Inutsuka's model (2005, 2006). }
   {At 1.2 mm we detected eight fragments towards this region, five located on its borders. The largest fragment has a mass of
   about 370 $M_{\odot}$. Class~I and Class~II young stellar objects are detected
   all over the region, with some observed far from the ionization front.
This result emphasises
   the possible importance of \emph{distant\/} interactions between the radiation, escaping from the ionized region, and the
   surrounding medium.}
   {}

\titlerunning{Triggered star formation on the borders of RCW~120}
   \maketitle
%
%________________________________________________________________

\section{Introduction \label{intro}}
The impact of massive stars on their environments can be either
destructive or constructive (Gorti \& Hollenbach \cite{gor02}; Tan
\& McKee \cite{tan01}). If constructive, a massive star will favour
subsequent star formation via energetic phenomena such as winds and
radiation, leading to a local increase in pressure. Indeed, many
physical processes can trigger star formation (see Elmegreen
\cite{elm98} for a review). The collect and collapse process is one
of them. In this process (first proposed by Elmegreen \& Lada
\cite{elm77}), the expansion of an \HII\ region generates the
formation of a layer of gas and dust that is accumulated between the
ionization front (IF) and the shock front (SF) that precedes the IF
on the neutral side. This compressed layer may become
gravitationally unstable along its surface and then fragment. This
process is interesting as it allows the formation of massive
fragments out of a previously uniform medium, thus permitting the
formation of massive objects, whether stars or clusters.

We have shown that the collect and collapse process has triggered
massive-star formation in Sh~104 (Deharveng et al.\ \cite{deh03})
and RCW~79 (Zavagno et al.\ \cite{zav06}); however, many questions
remain. In particular, the theoretical predictions are sometimes
difficult to match with the observations, since the `real'
physical environment (inhomogeneous medium, evidence of champagne
flow in some \HII\ regions, turbulence) is more complicated than
described in the models. Is the collect and collapse process an
efficient way of forming massive stars? How does star formation
proceed in the condensations formed via this process? Up to what
distance can a massive star have an impact on its surrounding?

We are engaged in a systematic study of a sample of \HII\ regions
selected on the basis of their simple morphology and their
potential for being collect and collapse regions (see Deharveng et
al.\ \cite{deh05} for details). These are choice locations for
studying the onset of massive-star formation in detail, and
RCW~120 is one of these regions.

In the present paper we examine the distribution of the cold dust
associated with RCW~120, determine the nature of the IR sources
observed towards it, and discuss the star formation processes
possibly at work in this region. Section~2 gives the distance,
identifies the exciting star and describes the morphology of
RCW~120. Section~3 presents new 1.2-mm continuum observations
giving an emission map over a $20\arcmin\times20\arcmin$ area.
Information derived from the 1.2-mm continuum emission (the cold
dust distribution and the mass of the observed fragments), along
with the properties of the YSOs observed in the direction of
RCW~120, are given in Sect.~4. The properties of YSOs are
discussed using the complete set of data available from the
large-scale IR surveys GLIMPSE (Benjamin et al.\ \cite{ben03}),
2MASS (Skrutskie et al.\ \cite{skr06}), and MSX (Egan et al.\
\cite{ega99}). Section~5 presents a discussion of the geometry of
RCW~120, the various star-forming processes identified towards
this region, and a comparison with the theoretical model of \HII\
region evolution of Hosokawa and Inutsuka (\cite{hos05}).
Conclusions are given in Sect.~6.

\section{Presentation of RCW~120 \label{present}}
\begin{figure}
   \centering
   \includegraphics[angle=0,width=90mm]{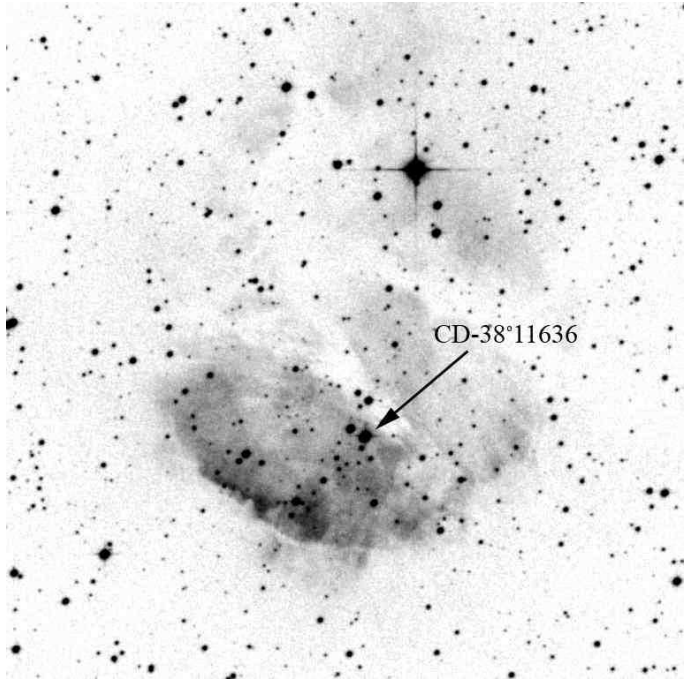}
   \caption{H$\alpha$ emission of RCW~120 taken from the DSS2-red survey.
The main exciting star, CD$-$38$\fdg$11636, is identified.}
              \label{Figha}
    \end{figure}

The \HII\ region RCW~120 (Rodgers, Campbell and
Whiteoak~\cite{rod60} -- also Sh2-3, Sharpless~\cite{sha59}) lies
in a direction close to the Galactic centre, slightly above the
Galactic plane, at $l=348\fdg249$, $b=0\fdg469$. RCW~120 is an
optical \HII\ region, of about
$7\arcmin\,$(E-W)$\times10\arcmin\,$(N-S) in size, located at
1.3~kpc (Russeil~\cite{rus03} and Sect.~\ref{dist}). It is excited
by an O8V star, LSS~3959, or CD$-$38$\fdg$11636
($\alpha_{2000}$=17$^{\rm h}$ 12$^{\rm m}$ 20.6$^{\rm s}$,
$\delta_{2000}$=$-$38$\degr$ 29$\arcmin$ 26$\arcsec$), identified
spectroscopically by Georgelin \& Georgelin~(\cite{geo70}). This
star is visible in Fig.~\ref{Figha}, taken from the DSS2-red
survey, showing the H$\alpha$ emission.

%-----------------------
    %\fig1
%-----------------------

%%%%%%%%%%%%%%
\subsection{Distance and exciting star \label{dist}}
%%%%%%%%%%%%%%

The molecular material associated with RCW~120 has a radial
velocity $V_{\rm LSR}$(CO) of $-8.7$~km~s$^{-1}$ (Blitz et
al.~\cite{bli82}). Using the Galactic rotation curve of Brand \&
Blitz~(\cite{bra93}) we derive a kinematic distance of 1.35~kpc.

The main exciting star of RCW~120 has been observed at various
wavelengths. Its magnitudes are $B=11.93$, $V=10.79$ (Avedisova \&
Kondratenko~\cite{ave84}), $J=8.013$, $H=7.708$, and $K=7.523$
(2MASS Point Sources Catalog [PSC]). Using the new calibrations of
Martins \& Plez~(\cite{mar06}), and the spectral type O8V, we
estimate a visual extinction A$_{\rm V}$ of $4.36$~mag and a
distance of 1.33~kpc. This photometric distance agrees with the
kinematic distance. We adopt a distance of 1.34~kpc in the
following.

RCW~120 is a thermal radio-continuum source. Its flux density has
been measured at various wavelengths, and is in the range
5.5--8.5~Jy (Manchester~\cite{man69}; Altenhoff et al.\
\cite{alt70}; Reifenstein et al.\ \cite{ref70}; Griffith et al.\
\cite{gri94}; Langston et al.\ \cite{lan00}). Adopting a flux
density of 7.0$\pm$1.5~Jy, a distance of 1.34~kpc, and using
Simpson \& Rubin's Eq.~1 (\cite{sim90}), we derive the ionizing
photon flux, $\log(N_{\rm{ Lyc}}$)=48.04$\pm$0.10. According to
the calibration of Martins et al.\ (\cite{mar05}), this
corresponds to a star of spectral type O8.5V--O9V. This is
somewhat later than the O8V spectral type estimated directly from
spectroscopy (Georgelin \& Georgelin~\cite{geo70}). But this is
not surprising as ionizing photons are very probably absorbed by
dust grains inside this \HII\ region; indeed, the emission of
these grains is observed at 21.3~$\mu$m (MSX Band E) clearly in
the direction of the ionized gas (see Fig.~1 of Deharveng et al.\
\cite{deh05}).

RCW~120 is notable for its high degree of symmetry. On the one
hand, the ionization front, traced by the 8\,$\mu$m emission, is
almost circular. On the other, the whole region presents a nearly
north-south symmetry axis, with the exciting star lying on this
axis.

Several facts indicate that a nearly north-south density gradient is
present in the region, with density increasing towards the south: i)
the zones of brightest H$\alpha$ emission are observed in the
southern part of the \HII\ region (see Figs.~\ref{Figha} and
\ref{FigPAH}); ii) the exciting star lies in the southern part of
the \HII\ region; and iii) the ionized gas is beginning to break out
of the \HII\ region, northwards. We come back to this point in the
next section and in Sect.~\ref{havi}.

%%%%%%%%%%%%%%%%%%
\subsection{Morphology of RCW~120}
%%%%%%%%%%%%%%%%%%
\begin{figure*}
   \centering
   \includegraphics[angle=0,width=120mm]{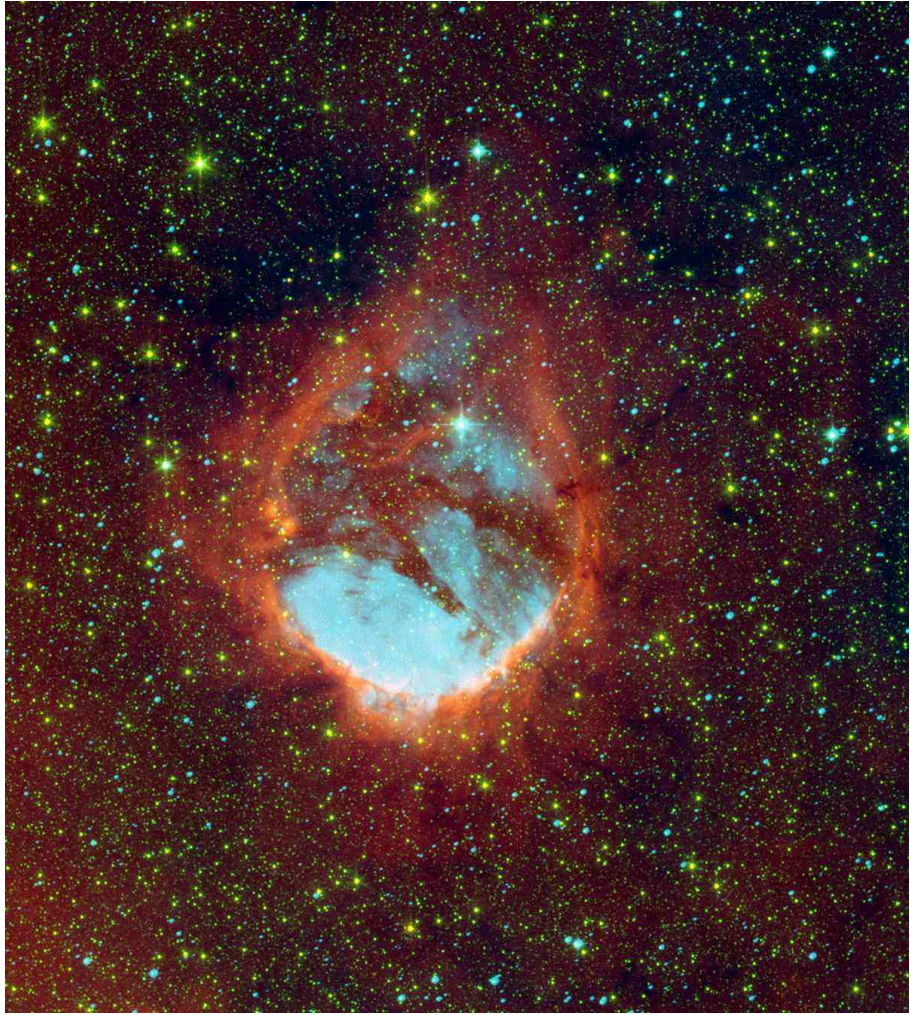}
   \caption{Spitzer-IRAC 4.5\,$\mu$m (green) and 8\,$\mu$m (red) images from the GLIMPSE survey
superimposed on a SuperCOSMOS H$\alpha$ image (turquoise) of
RCW~120. The 8\,$\mu$m emission is dominated by the polycyclic
aromatic hydrocarbon (PAH) molecules. Note the presence of a PAH
layer, corresponding to the hot photodissociation region that
surrounds the ionized region. The field size is 24\arcmin\, (E-W)
$\times$ 27\arcmin\, (N-S). North is up and east is left}
   \label{FigPAH}%
\end{figure*}
%fig2
RCW~120 is associated with a large amount of dust.
Figure~\ref{FigPAH} is a composite colour image showing the
H$\alpha$ emission (taken from the Super-COSMOS survey, Parker et
al.\ \cite{par05}) and the 4.5\,$\mu$m and 8\,$\mu$m emission (from
the Spitzer-GLIMPSE survey, http://www.astro.wisc.edu/sirtf/). The
8\,$\mu$m IRAC band contains emission bands centred at 7.6, 7.8, and
8.6\,$\mu$m, commonly attributed to polycyclic aromatic hydrocarbon
(PAH) molecules and a continuum contribution from very small grains.
PAHs are believed to be destroyed in the ionized gas, but larger
molecules can survive (Verstraete et al.~\cite{ver96}; Peeters et
al.~\cite{pee05}). PAHs are excited by UV photons ($h\nu < $13.6~eV)
(Sellgren~\cite{sel84}) in the photo-dissociation region (PDR). PAH
emission is a good tracer of the hot PDR that surrounds the \HII\
region. Figure~\ref{FigPAH} shows that PAH emission completely
surrounds the ionized gas. Filaments extend far from the ionization
front, suggesting the idea of a `leaky' \HII\ region, probably due
to small-scale inhomogeneities in the ionization front and in the
surrounding medium. This point is important, as it could explain how
triggered star formation may occur far from the ionization front
(see Sect.~\ref{farsf}).
\begin{figure*}
   \centering
   \includegraphics[angle=0,width=180mm]{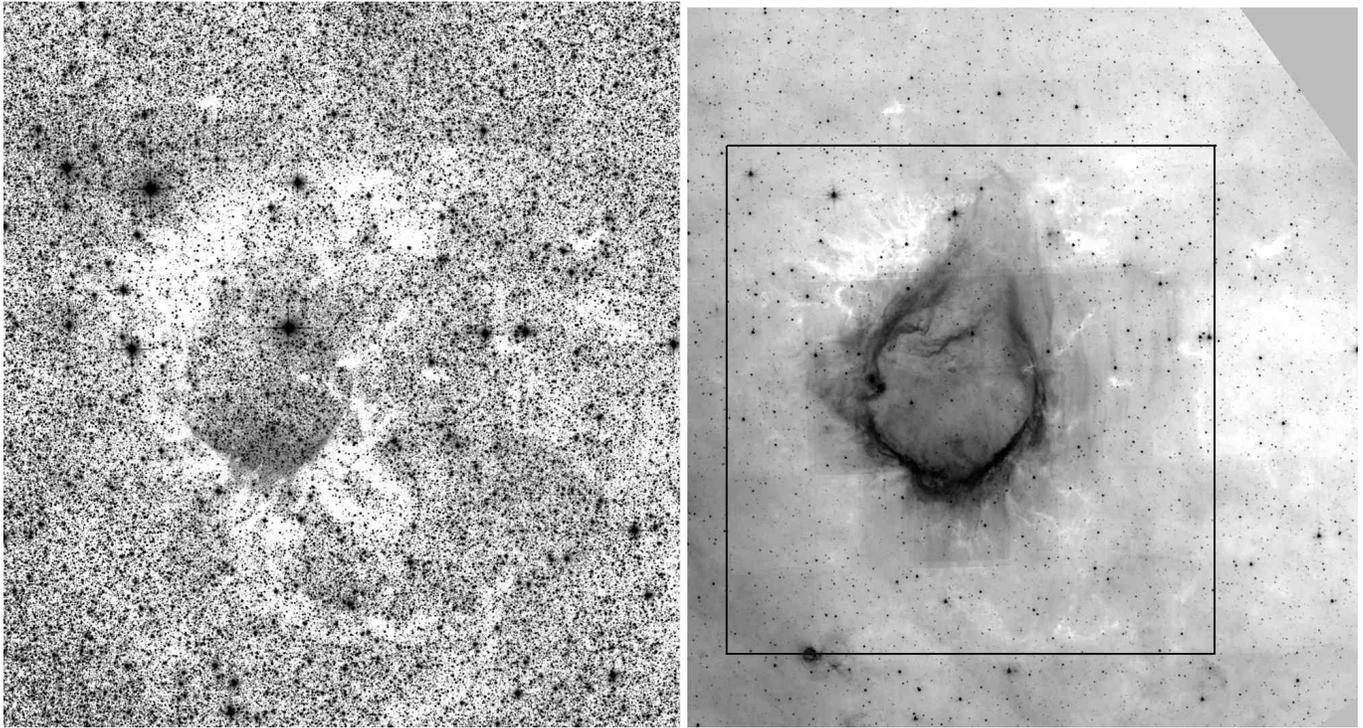}
   \caption{Dust associated with RCW~120. {\it Left:} $K_{\rm S}$ (2.17\,$\mu$m) mosaic image
   of the RCW~120 region from the 2MASS survey. Zones of high absorption surround the
   ionized region. Absorption filaments extend far away from the ionization front, revealing a non-homogenous medium. {\it Right:}
   PAH emission at 8\,$\mu$m from Spitzer-GLIMPSE. Absorption features are still seen at this wavelength, indicating the presence of
   dense material. The black box represents the region that has been searched for young
stellar objects }
   \label{Figabs}%
\end{figure*}
%fig3
Figure~\ref{Figabs} (left) is a mosaic $K_{\rm S}$ image of
RCW~120 from the 2MASS survey. This clearly shows the presence of
an absorbing region surrounding RCW~120, with some filamentary
structures that extend far from the ionized region. The clear
limits of the absorbing zone and its large width (especially to
the north) suggest that we are seeing the remaining part of the
parental molecular cloud. The right part of Fig.~\ref{Figabs}
shows, on the same scale, an image of the RCW~120 region, taken
from the Spitzer-GLIMPSE survey in the band centred at 8\,$\mu$m.
The absorption zones, clearly visible at 2\,$\mu$m, are still
observed at 8\,$\mu$m (especially to the north), indicating the
presence of high-density material in those regions. Lacking
velocity information, we cannot prove the direct association of
the absorbing zones with the ionized region. The high contrast
seen at 2\,$\mu$m (Fig.~\ref{Figabs} left), between the absorbing
zones that closely surround RCW~120 and the surrounding medium
observed at a larger distance, indicates that the association may
be real. The black box shown in Fig.~\ref{Figabs} on the 8\,$\mu$m
GLIMPSE image shows the zone that we have selected to search for
YSOs using colour criteria in the near and mid IR. This part is
described in Sect.~\ref{yso}.

\section{Observations \label{obs}}
%\subsection{SEST-SIMBA 1.2-millimetre continuum imaging \label{mmobs}}
Of the two sets of observations, the SEST-SIMBA contains continuum
maps at 1.2-mm (250~GHz) of a 20\arcmin $\times$ 20\arcmin\ field
centred on RCW~120 were obtained using the 37-channel SIMBA
bolometer array (SEST Imaging Bolometer Array) on May 7-8, 2003. The
beam size is 24\arcsec. The positional uncertainty of the SIMBA
observations is less than 3$\arcsec$. Six individual maps covering
the whole region were obtained with the fast scanning speed
(80\arcsec\ per second). The total integration time was 10 hours.
The final map was constructed as described in Zavagno et al.\
(\cite{zav06}). The residual noise in the final map is about
20~mJy/beam (1$\sigma$). We then used the emission above 5$\sigma$
(100 mJy/beam level) to delineate the 1.2-mm condensations.
%\subsection{H$\alpha$ observations}

The H$\alpha$ Fabry-Perot observations of RCW~120 were obtained with
CIGALE on a 36-cm telescope (La Silla, ESO). CIGALE uses a
Fabry-Perot interferometer scanning the H$\alpha$ profile to give
the kinematics over the field. The field of view is 39$\arcmin$
square with a pixel size of 9$\arcsec$. The Fabry-Perot
interferometer has an interference order of 2604 at H$\alpha$, i.e.\
a free spectral range of 115~km~s$^{-1}$. The finesse is 10 (FWHM
11.5~km~s$^{-1}$) and the sampling step is 5~km~s$^{-1}$. The
velocity and FWHM uncertainties are both $\sim$1~km~s$^{-1}$. A
complete description of the instrument, including data acquisition
and reduction techniques, can be found in Le Coarer et
al.~(\cite{lec92}).

The H$\alpha$ profiles need to be decomposed into two components:
the night-sky lines (geocoronal H$\alpha$ and OH) and the nebular
lines. The night-sky lines are modelled by the instrumental
profile, while the nebular lines are modelled by Gaussians
convolved with that profile. In order to increase the
signal-to-noise ratio we extracted and analysed profiles from
large areas ($45\arcsec \times 45\arcsec$). The profile analysis
shows that the LSR radial velocity of the ionized gas of RCW~120
ranges from $-$8~km~s$^{-1}$ to $-$15~km~s$^{-1}$.
\section{Results}
%\subsection{Continuum imaging at 1.2~mm \label{mmres}}
%____________________________________________________
%
\begin{figure*}[tb]
 \includegraphics[angle=0,width=180mm ]{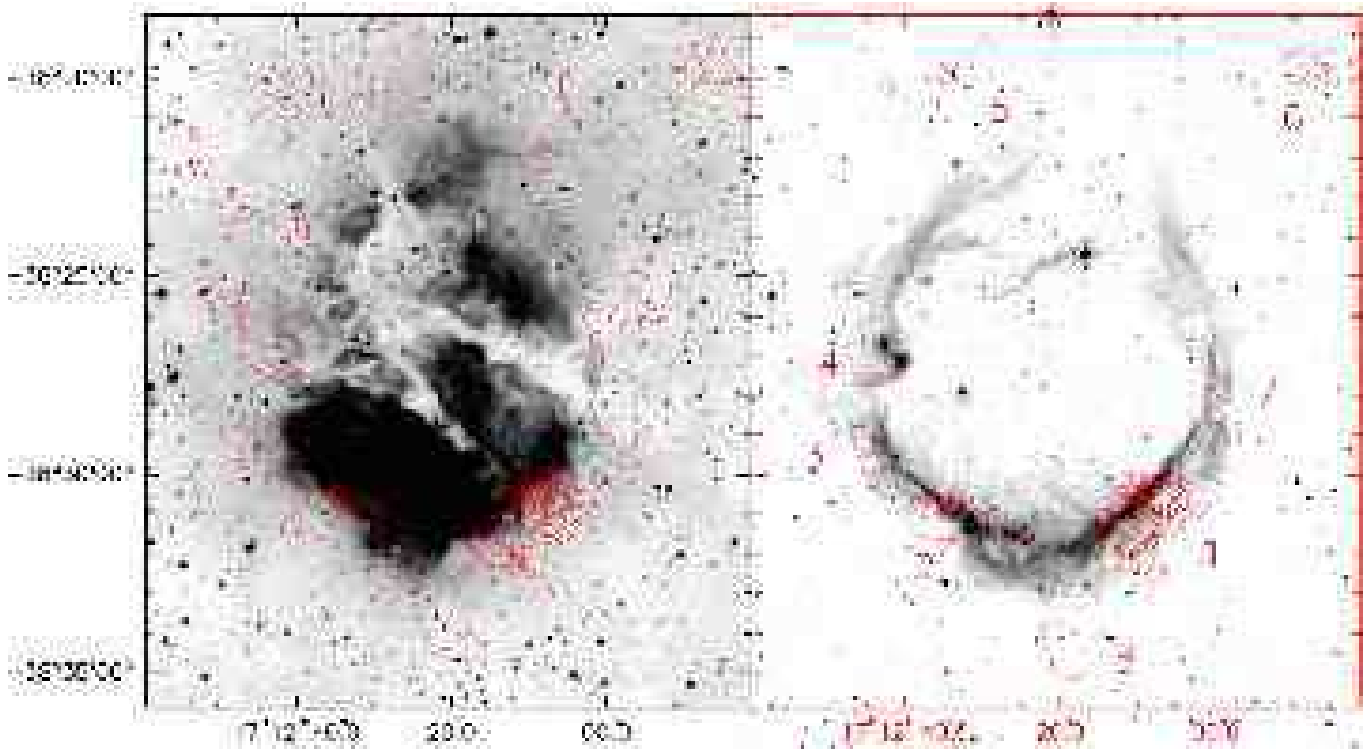}
  \caption{{\it Left}: Millimetre continuum emission contours superimposed on a SuperCOSMOS H$\alpha$ image of the
  region. Contours range from 60 (3$\sigma$) to 960~mJy/beam in steps of
100~mJy/beam
  and from 1000 to 3000~mJy/beam in steps of 400~mJy/beam. {\it Right}: Millimetre continuum emission contours superimposed on the GLIMPSE 5.8\,$\mu$m
  image of the region. Contours range from 100~mJy/beam (5$\sigma$) to 400 in steps of 100~mJy/beam and then from 500
  to 3000 in steps of 250~mJy/beam. Condensations at 1.2~mm are identified in Table~1}
  \label{mm}
\end{figure*}
%fig4
Figure~\ref{mm} (left) shows the 1.2-mm continuum emission contours
superimposed on a SuperCOSMOS H$\alpha$ image of RCW~120. Eight
condensations are observed at a 5$\sigma$ level. Five of these (nos\
1, 2, 3, 4, and 7) are immediately adjacent to the ionized region.
This location suggests that these condensations may result from the
fragmentation of a layer of collected material accumulated between
the ionization front and the shock front, as predicted by the
collect and collapse model. The two most massive condensations (nos
1 and 2) are observed to the south. Condensations 5, 6, and 8 are
observed farther away from the ionization front but coincide with
regions of high extinction in Fig.~\ref{Figabs}. Note that only low
extinction is detected in the optical and 8\,$\mu$m images in the
direction of the most massive condensations (1 and 2), indicating
that most of the absorbing material must be located {\it behind} the
emitting region.

Figure~\ref{mm} (right) presents the 1.2-mm continuum emission
contours superimposed on the 5.8\,$\mu$m GLIMPSE image of RCW~120.
The lowest contour delineates the condensations' surfaces as
defined, for the mass estimates, at the 5$\sigma$ level
(100~mJy/beam, Sect.~\ref{obs}).
\subsection{Mass estimates\label{massest}}
The millimetre continuum emission from the condensations identified
in Fig.~\ref{mm} (right) is mainly due to optically thin thermal
dust emission. We used the formula given by Hildebrand
(\cite{hil83}) and proceeded as explained in Zavagno et
al.~(\cite{zav06}). Table~\ref{massemm} lists the measured and
derived properties obtained for the millimetre fragments identified
in Fig.~\ref{mm}. Column 1 gives the fragment numbers, columns 2 and
3 give the emission peak coordinates, column 4 gives the 1.2-mm
integrated flux, and column~5 the range of derived masses for the
corresponding fragment depending on the adopted temperature (20 or
30~K). The lower mass values correspond to the higher dust
temperature. Because of possible molecular line contamination
(overestimation of the continuum flux) and the possible presence of
an internal source of heating (a higher dust temperature than
expected), the fragments' mass estimate is an upper limit (see also
Beuther et al.\ \cite{beu02}). If an outflow is present, the
molecular line contamination to the measured 1.2-mm flux may be as
much as 30\% (Gueth et al. \cite{gue03}) but is probably about 10\%
if no outflow is present (Guilloteau, private communication).
\begin{table*}
\caption{Mass estimates for the millimetre fragments}
\begin{tabular}{c l l r r}

\hline\hline
 Number & \multicolumn{2}{c}{Peak position} & $F_{\mathrm{1.2mm}}$$^{\footnotesize 1}$ & Mass range$^{\footnotesize 2}$ \\
        &            \multicolumn{1}{c}{$\alpha_{2000}$} & \multicolumn{1}{c}{$\delta_{2000}$}   & (mJy)              & \multicolumn{1}{c}{($M_{\odot}$)} \\
  \hline
        &            &     &                  &                   \\
  1 & 17$^{\rm h}$ 12$^{\rm m}$ 08$\fs$94 &  $-$38$\degr$ 30$\arcmin$ 43$\farcs$20  & 16100 & 278 -- 465    \\
  2 & 17$^{\rm h}$ 12$^{\rm m}$ 34$\fs$16 &  $-$38$\degr$ 30$\arcmin$ 51$\farcs$95 & 4100 & 71 -- 118    \\
  3 & 17$^{\rm h}$ 12$^{\rm m}$ 44$\fs$78 &  $-$38$\degr$ 29$\arcmin$ 28$\farcs$30 & 540 & 9 -- 15 \\
  4 & 17$^{\rm h}$ 12$^{\rm m}$ 41$\fs$50 &  $-$38$\degr$ 27$\arcmin$ 07$\farcs$43 & 797   & 13 -- 23 \\
  5 & 17$^{\rm h}$ 12$^{\rm m}$ 32$\fs$06 &  $-$38$\degr$ 19$\arcmin$ 48$\farcs$30  & 323 & 5 -- 9 \\
  6 & 17$^{\rm h}$ 11$^{\rm m}$ 47$\fs$90 &  $-$38$\degr$ 19$\arcmin$ 48$\farcs$28  & 1003 & 18 -- 30 \\
  7 & 17$^{\rm h}$ 11$^{\rm m}$ 58$\fs$72 &  $-$38$\degr$ 28$\arcmin$ 27$\farcs$18   & 422 & 7 -- 12  \\
  8 & 17$^{\rm h}$ 12$^{\rm m}$ 19$\fs$82 &  $-$38$\degr$ 34$\arcmin$ 04$\farcs$14   & 984 & 17 -- 28 \\
  \hline
  \label{massemm}
\end{tabular}\\
\\
{$^{\footnotesize 1}$ 1.2-mm flux integrated above the 5$\sigma$ level} \\
{$^{\footnotesize 2}$ The lower mass is calculated for $T_{\rm{dust}}$=30~K, the higher mass for $T_{\rm{dust}}$=20~K \\
}
\end{table*}
The two most massive fragments, nos 1 and 2, are located to the
south, probably a consequence of the higher density in this area.
The other fragments, located on the borders of the ionized region or
even farther away, are less massive. Fragment~1 shows some
structure, in particular a highly peaked emission. The 1.2-mm
emission peak of 3.08~Jy/beam indicates an H$_2$ column density
greater than 3.5~$\times 10^{23}$ cm$^{-2}$ for a temperature of
20~K, corresponding to a visual extinction $A_V$$\geq$ 200~mag. This
fragment is a potential site of high-mass star formation (see also
Beuther et al.\ \cite{beu02}). However, as discussed in
Sect.~\ref{yso}, no 8\,$\mu$m source is observed towards this
emission peak. This could be due to the low sensitivity of the
GLIMPSE survey. A source should be sought at longer wavelengths.
\subsection{Search for young stellar objects using 2MASS and GLIMPSE data \label{yso} }
Our purpose is to look for star formation towards RCW~120. For
this we use the Spitzer-GLIMPSE survey to do a systematic search
for YSOs using colour selection criteria. Indeed, YSOs are
expected to have specific colours depending on mass and
evolutionary status (see Allen et al.\ \cite{all04}). We have
selected the absorbing zone, centred on RCW~120, seen in the near
IR (Fig.~\ref{Figabs}), from $\alpha_{2000}$ from 17$^{\rm h}$
11$^{\rm m}$ 11$\fs$05 to 17$^{\rm h}$ 13$^{\rm m}$ 17$\fs$54 and
from $\delta_{2000}$ from $-$38$\degr$ 14$\arcmin$ 30$\farcs$40 to
$-$38$\degr$ 40$\arcmin$ 40$\arcsec$.

We extracted the 35178 objects in this zone from the GLIMPSE PSC
(http://www.astro.wisc.edu/sirtf/). Then, to avoid sources near
the detection limit, we selected the 2654 sources that had been
measured in the four IRAC bands and were brighter than 11~mag in
the 8\,$\mu$m band. From these we selected those objects with
colours corresponding to Class~II and Class~I, i.e.\ $[3.6]-[4.5]
\geq 0.4$ and $[5.8]-[8.0] \geq 0.4$ (Allen et al.\ \cite{all04}).
The final selection is listed in Table~2 to which we have added
the exciting star of RCW~120 and a few giants.

Table~2, available at the CDS (http://cdsweb.u-strasbg.fr/A+A.htx),
gives the position and photometry from 1.25\,$\mu$m to 8\,$\mu$m of
the detected YSOs, sorted by location (towards the millimetre
condensations, towards and outside from the \HII\ region). Column~1
gives their identification numbers. Columns~2 and 3 give their
coordinates according to the GLIMPSE PSC. Columns 4 to 6 give their
$J$, $H$, and $K_{\rm S}$ magnitudes from the 2MASS PSC
(http://tdc-www.harvard.edu/software/catalogs/tmpsc.html). Columns 7
to 10 give their [3.6], [4.5], [5.8], and [8.0] magnitudes from the
GLIMPSE PSC. When not available from the PSC, we measured the
magnitudes (aperture photometry) using the Basic Calibrated Data
frames; these are indicated with asterisks in Table~2. Column~11
gives general comments about the nature of each source (Class~I,
Class~II, giant). \addtocounter{table}{1}

\subsection{Properties of the observed sources}
Several tools can be used to determine the nature of the sources,
using their IR magnitudes from the 2MASS, Spitzer-GLIMPSE, and MSX
surveys.

\begin{figure*}
\includegraphics[angle=0,width=120mm ]{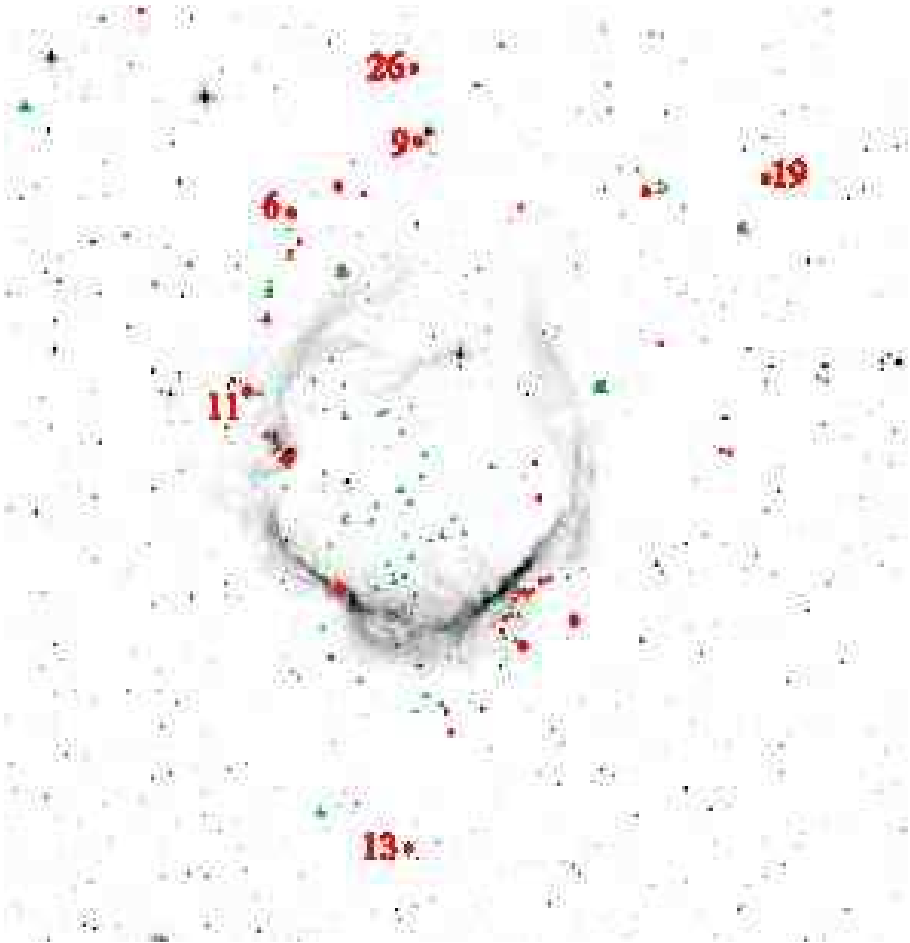}
  \caption{Spatial distribution of the YSOs detected towards
RCW~120. The red circles are Class I sources, the green triangles
Class II  sources. In both cases, the largest symbols are for the
brightest sources, with $[8.0]\leq$6~mag. Bright Class~I sources
observed far from the ionization front are identified (see
Sect.~5.2). The field size is 25\arcmin\, (E-W) $\times$ 25\farcm5\,
(N-S)}
  \label{spacedist}
\end{figure*}

\begin{figure*}
\includegraphics[angle=270,width=120mm ]{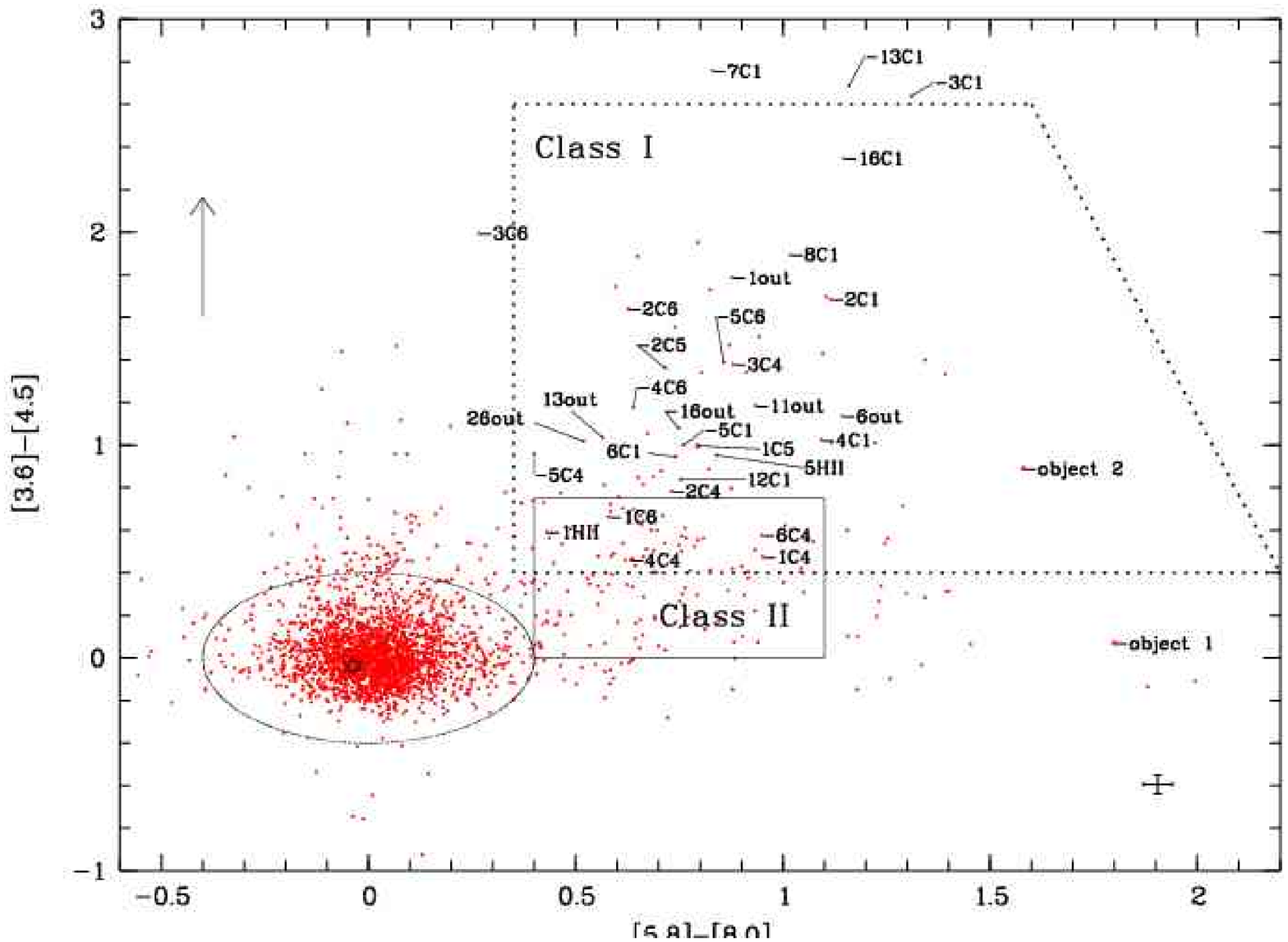}
  \caption{GLIMPSE colour-colour diagram, $[3.6]-[4.5]$ versus
$[5.8]-[8.0]$, for the sources observed towards RCW~120. Class~I and
Class~II zones are indicated according to criteria given by Allen et
al. (2004). The black arrow is the reddening vector for a visual
extinction of 40~mag. The ellipse centred on 0,0 encloses the region
of main sequence and giant stars. The black star represents the
exciting star of RCW~120 }
  \label{ccspitzer}
\end{figure*}

Figure~\ref{spacedist} shows the spatial distribution of the
Class~I and Class~II YSOs identified in the direction of RCW~120
and shows that these YSOs are located in three main zones: i) near
the cold dust condensations identified in Fig.~\ref{mm}; ii) in
the direction of the ionized region; and iii) far from the
ionization front and, in some cases, not associated with a
detected millimetre condensation. The lack of velocity information
makes it impossible to firmly associate the detected YSOs with
RCW~120. However, the statistical analysis presented in
Sect.~\ref{farsf} suggests that most of these objects are
associated with this region.
%fig5
The GLIMPSE colour-colour diagram is shown in
Fig.~\ref{ccspitzer}. The sources discussed in the text are
identified by the labels given in Table~2.
%fig6
A large number of objects are Class~I sources. We do not consider
those sources located in the lower part of the Class~II box, since
main sequence and giant stars (located in the elliptical region
centred on 0,0 in this $[3.6]-[4.5]$ versus $[5.8]-[8.0]$ diagram)
may be falsely displaced into this zone if they are faint and
superimposed on background emission (the colours of the extended
emission are $[3.6]-[4.5]=0.1$ and $[5.8]-[8.0]=1.9$). Giant stars
can be bright IR sources. We used the 2MASS $J-H$ versus $H-K$
diagram, not presented here, to identify such stars.

Note that the colour-colour diagrams do not give information about
the sources' brightnesses and that additional information is
needed to better discuss the nature of the sources. We do not
discuss the spectral energy distribution (SED) in this paper,
because the 2MASS data for the embedded sources are often missing
(or are given as lower limits) and also because the wavelength
coverage of GLIMPSE is too limited to accurately fit the SEDs,
especially since GLIMPSE sees only part of the deep silicate
absorption that is often present in YSOs. Longer wavelength data
are needed to use the SED as a reliable tool for evolutionary and
brightness classifications.

Below we give information about some specific properties of the
sources, due either to their locations and/or evolutionary stages.
Sources observed towards millimetre condensations may be affected by
local extinction. Apparently, Class~I sources may in fact be
reddened Class~II. This is particularly critical for deeply embedded
sources classified as Class~I with no 2MASS counterparts.

\begin{figure}
\includegraphics[angle=0,width=85mm ]{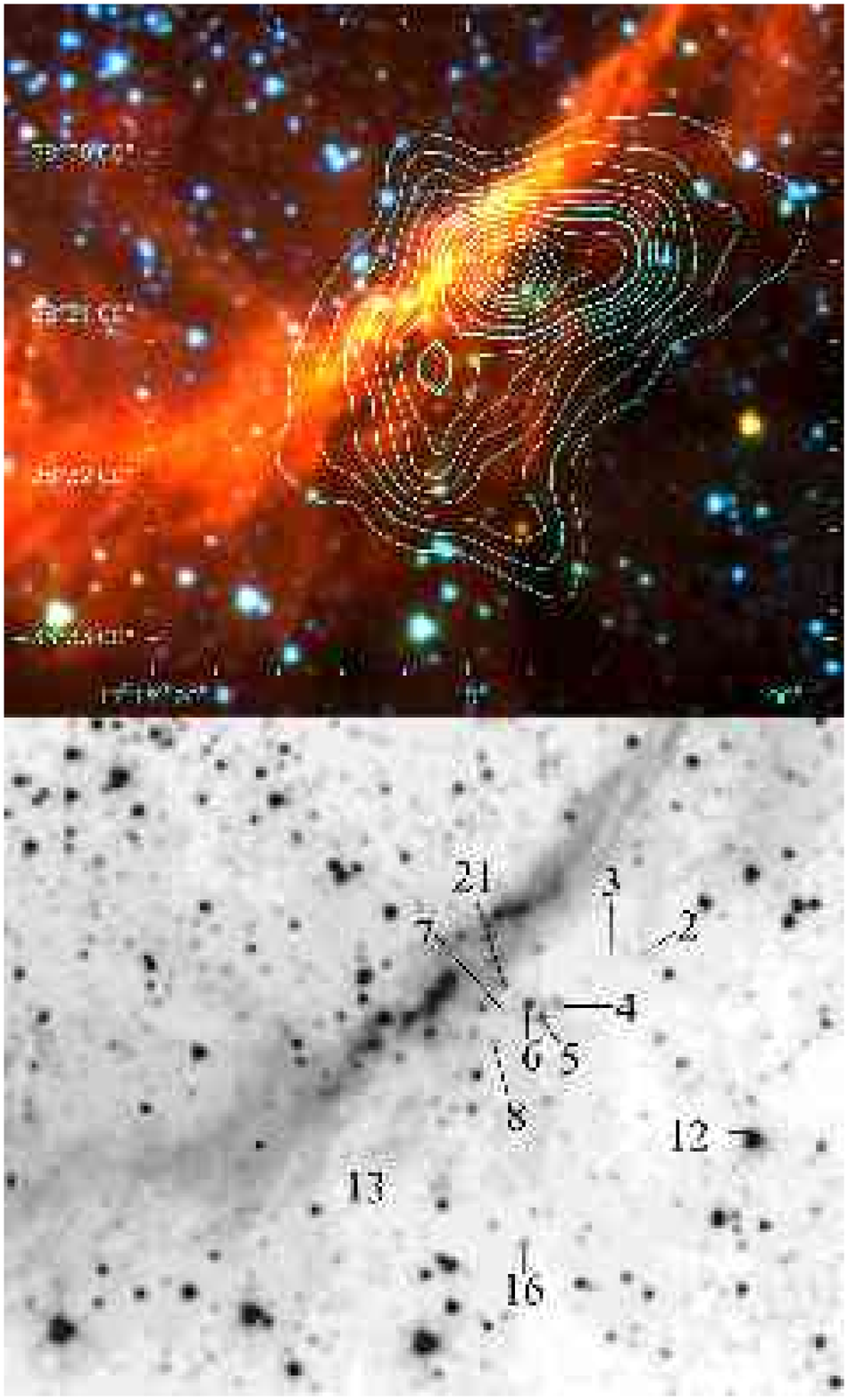}
  \caption{{\it Top}: colour composite image of condensation~1, showing the $K$ image (blue) from
  2MASS, and
  the 3.6\,$\mu$m (green) and 8\,$\mu$m (red) image from the
  GLIMPSE survey. {\it Bottom}: The sources discussed in the text are identified on the 5.8\,$\mu$m GLIMPSE image}
  \label{spitc1}
\end{figure}

\noindent{\bf{Condensation 1}} \\
Figure~\ref{spitc1} is a composite colour image of condensation 1.
%fig7
Source 12C1, the brightest object at 8\,$\mu$m near condensation 1,
is located in the transition region between Class~I and Class~II
sources (Fig.~\ref{ccspitzer}). This source is not directly
associated with 1.2-mm emission but is observed towards a
filamentary structure observed in absorption at 8\,$\mu$m (see
Fig.~\ref{Figabs}); it is far (about 1.2~pc) from the ionization
front but is observed surrounded by diffuse 8\,$\mu$m emission,
indicating that far UV photons leaking from the ionized region have
reached this zone. This radiation may have influenced the formation
of this young source. We discuss this point in Sect.~\ref{farsf}.

Source 16C1, a Class~I object, is the brightest source observed
towards condensation 1, and it has no 2MASS counterpart. This source
coincides with an extension of the 1.2-mm continuum emission and is
observed at the head of a finger-shaped absorbing region observed at
8\,$\mu$m. Faint 8\,$\mu$m emission surrounds this region,
indicating that UV photons reach this zone. The location of this
source at the vertex of an absorbing structure indicates that
globule squeezing may have occurred here, triggering star formation.

Source 13C1 is also a Class~I object having no 2MASS counterpart.
It is observed in the direction of a 1.2-mm emission extension.

The sources 4C1, 5C1, 6C1, 7C1, 8C1, and 21C1 are observed towards
the main 1.2-mm emission peak. Most of these are low-luminosity
Class~I objects. This strong emission peak does not coincide with an
absorbing zone at 8\,$\mu$m, so the condensation must be located
behind the 8\,$\mu$m emitting region. The nature of source 21C1,
observed exactly coincident with the 1.2-mm peak, is unclear, as
this source has no measured magnitudes in the [3.6], [5.8], and
[8.0] bands.

\begin{figure}
\includegraphics[angle=0,width=85mm ]{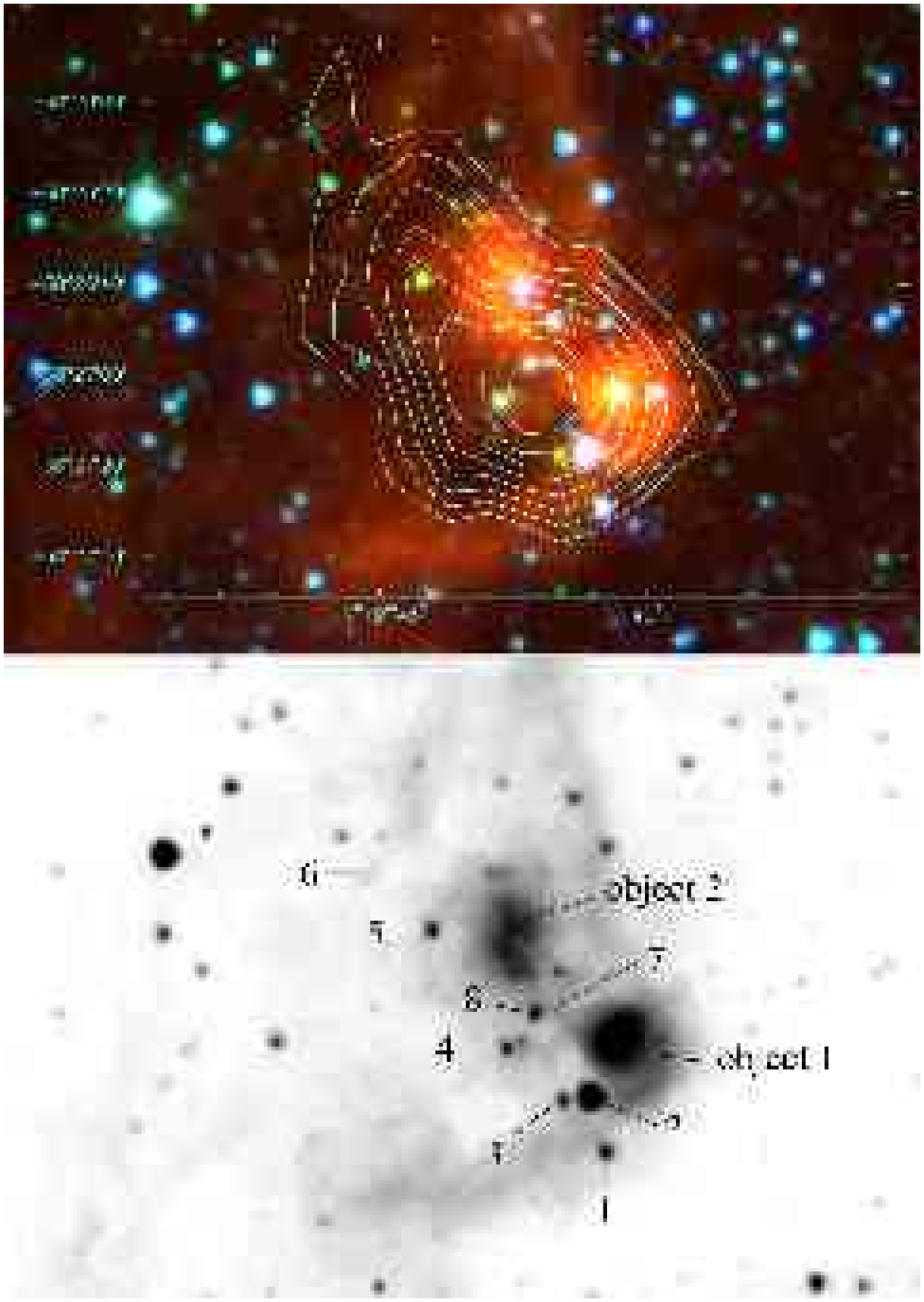}
  \caption{Top: Same as Fig.~\ref{spitc1}, but for condensation 4. The sources discussed in the text are
  identified in the 8\,$\mu$m GLIMPSE image (bottom)}
  \label{spitc4}
\end{figure}

\noindent{\bf Condensation 4} \\
Figure~\ref{spitc4} shows a detailed view of condensation 4.
%fig8
As seen in Fig.~\ref{spitc4}, two bright extended sources called
objects~1 and 2 and not given in the GLIMPSE PSC (probably due to
their extended nature), dominate the 8\,$\mu$m emission. At their
centres lie near-IR stellar objects. Objects~1 and 2 are also seen
in the MSX images and are classified from their MSX colours as
Herbig Ae/Be objects (for object~1 see source 19 in Deharveng et
al.\ \cite{deh05}). We have verified from its MSX colours that
object~2 (not included in Deharveng et al. \cite{deh05}) is a Herbig
Ae/Be object as well. The extended nebulae are probably local PDRs
created by the radiation of the central sources. These are not
massive enough to form \HII\ regions but they can, with lower energy
photons, heat the surrounding dust, thus creating local PDRs.

Class~I sources including nos 2, 3, and 4C4, are observed in the
immediate surroundings, in addition to these two Herbig Ae/Be
stars. No source is detected towards the peak of the 1.2-mm
condensation. Note that the shape of the ionization front is
distorted in the direction of condensation~4. We discuss this
point in Sect.~\ref{farsf}.

\noindent{\bf Condensation 5} \\
Condensation 5 is observed far from the ionization front and
coincides with a region of high extinction observed at both
2\,$\mu$m and 8\,$\mu$m (Fig.~\ref{Figabs}). An interesting point is
the presence of numerous {\it radial} structures seen in absorption
at 8\,$\mu$m, perpendicular to the IF (Fig.~\ref{Figabs}). This
suggests that radiation passes through this rather dense medium,
with enough energy to shape it, and may have favoured subsequent
star formation within this possible pre-existing condensation.

Two Class~I sources, 1C5 and 2C5, are observed towards
condensation 5. The bright source 1C5 is observed towards the
condensation's centres.

\begin{figure}
\includegraphics[angle=0,width=85mm ]{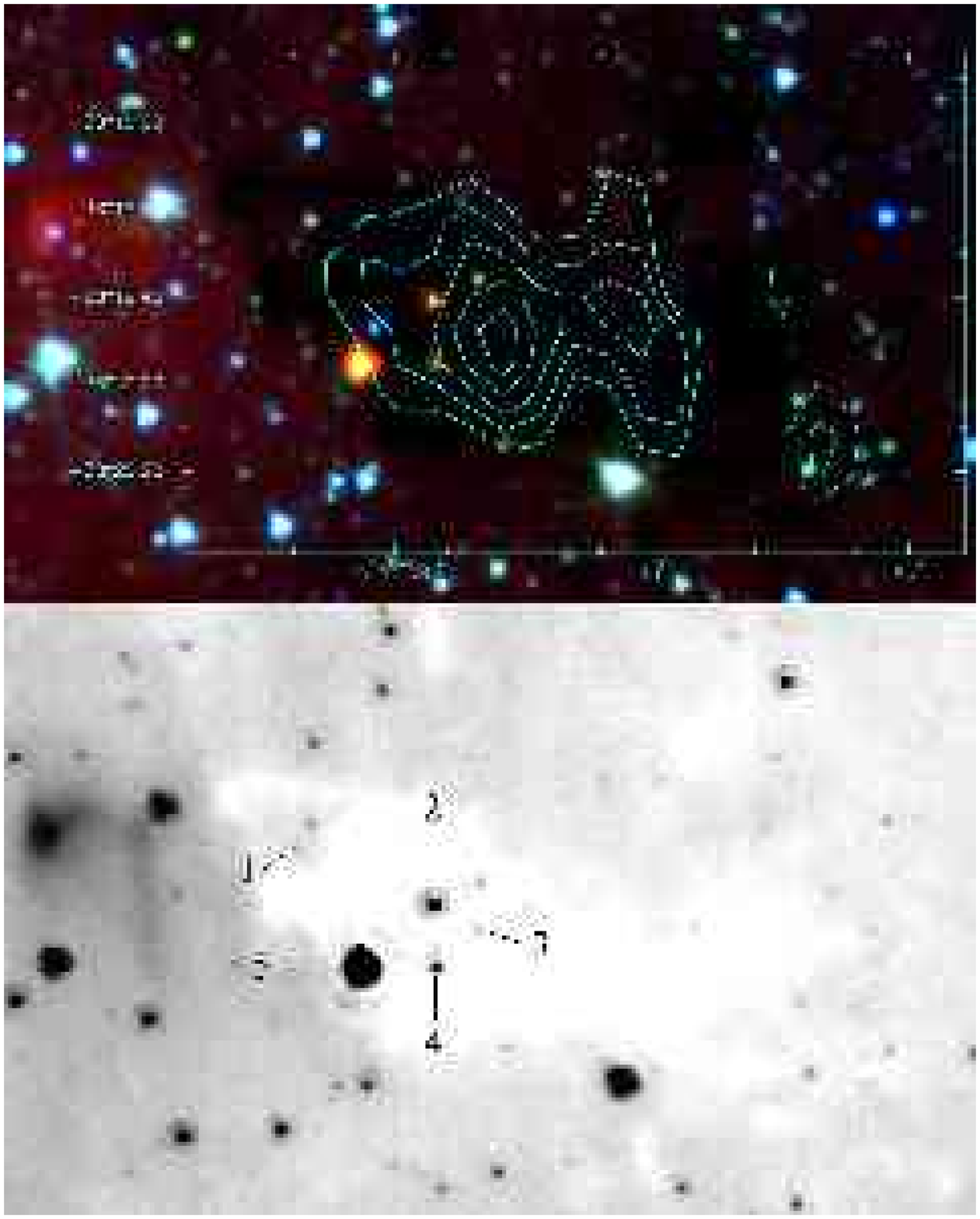}
  \caption{Same as Fig.~\ref{spitc1} but for condensation 6. The sources discussed in the text
  are identified in the 5.8\,$\mu$m GLIMPSE image (bottom)}
  \label{spitc6}
\end{figure}

\noindent{\bf Condensation 6} \\
Figure~\ref{spitc6} shows condensation 6 in detail.  This object,
like condensations 5 and 8, is observed far from the ionization
front (about 1.5~pc away) and coincides with an absorption region
seen in the 2\,$\mu$m and 8\,$\mu$m images (Fig.~\ref{Figabs}).
%fig9
Five sources are observed towards condensation 6, source 5C6 being
the brightest. Source 3C6 is a very faint red object observed
towards the peak of the 1.2-mm emission (Fig.~\ref{spitc6}). Objects
2C6 and 4C6, also classified as Class~I, are located farther from
the centre of the 1.2-mm emission peak.

\noindent{\bf{Sources observed towards the ionized region}} \\
As seen in Table~2 and in Fig.~\ref{spacedist}, all of the sources
observed towards the ionized region are Class~II objects, apart
from one Class~I source, 13\HII. The ionizing star of RCW~120 is
identified in Fig.~\ref{Figha}. Its GLIMPSE colours correspond to
those of main sequence stars (Fig.~\ref{ccspitzer}).

A Class~I source (5\HII) and a Class~II source (1\HII) are
observed towards the ionized region, lying on the vertices of
triangular structures. Both (the sources and the structures) are
seen in emission at 8\,$\mu$m (see Fig.~\ref{inst}) and are
discussed in Sect.~\ref{sf}.

\subsection{H$\alpha$ velocity field \label{havi}}
\begin{figure}
 \includegraphics[angle=0,width=85mm ]{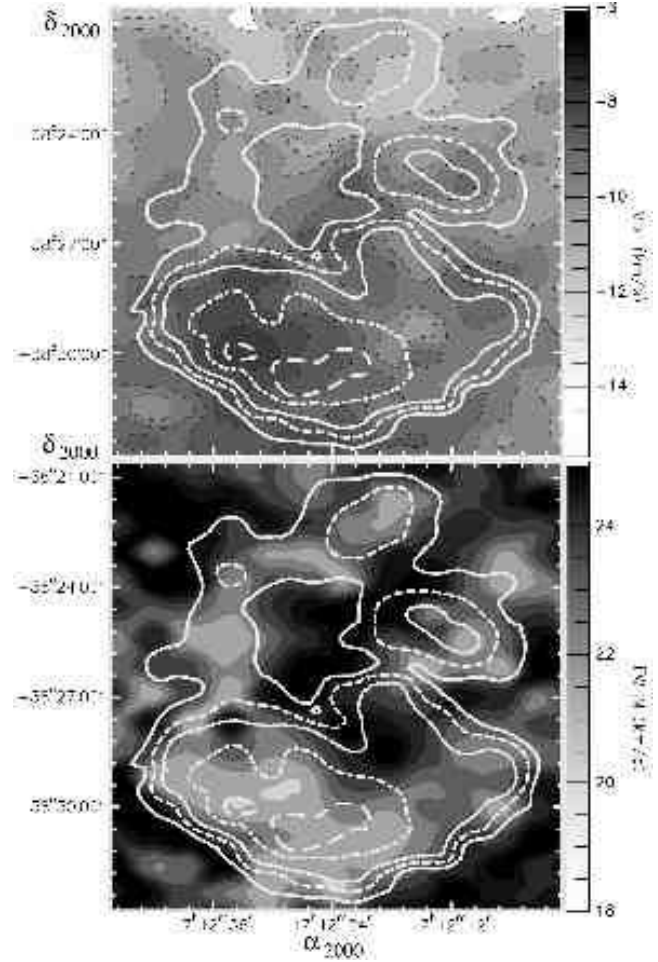}
  \caption{{\it Top}: H$\alpha$ velocity field (grey scale plus black dashed contours). {\it Bottom}: H$\alpha$ line width (grey scale).
  The H$\alpha$ emission is superimposed as white contours }
  \label{havit}
\end{figure}

Figure~\ref{havit} presents the velocity field of the ionized gas
observed towards RCW~120 and the width of the H$\alpha$ emission
line over the field. The uncertainty of the H$\alpha$ line width is
1~km s$^{-1}$.
%fig10
The velocity field displays a gradient, from about $-$9~km
s$^{-1}$ to the south of the \HII\ region, up to $-$16~km~s$^{-1}$
to the north of the region. In Fig.~\ref{FigPAH} RCW~120 looks
like a bottle full of ionized gas, its `neck' turned to the north
where a clear opening in the 8\,$\mu$m indicates a break in the
ionization front and the surrounding dust layer. The H$\alpha$
velocity field shows that the ionized gas flows towards the
observer. This result is consistent with the most massive
condensation being located to the south where dense material has
accumulated. The northern part is probably less dense, allowing
the ionized gas to break the surrounding shell and flow away. We
are probably observing the beginning of a `champagne' phase
(Tenorio-Tagle \cite{ten79}), and the observed shape of the region
is also consistent with this result.

Figure~\ref{havit} shows that on the borders of the ionized region
the H$\alpha$ line width is narrow -- around 18~km s$^{-1}$ -- when
compared to the central parts where the width is $\simeq$24~km
s$^{-1}$. This is expected for an expanding \HII\ region: whereas
the expansion velocity has no radial component on the borders of
this region (thus the line is narrow), the expansion velocity is
purely radial  in the direction of the centre and we see both the
approaching and receding sides (thus broadening the line). We
expected this to be confirmed by the \HI\ emission but it was not,
as described below.

\begin{figure}
 \includegraphics[angle=0,width=85mm ]{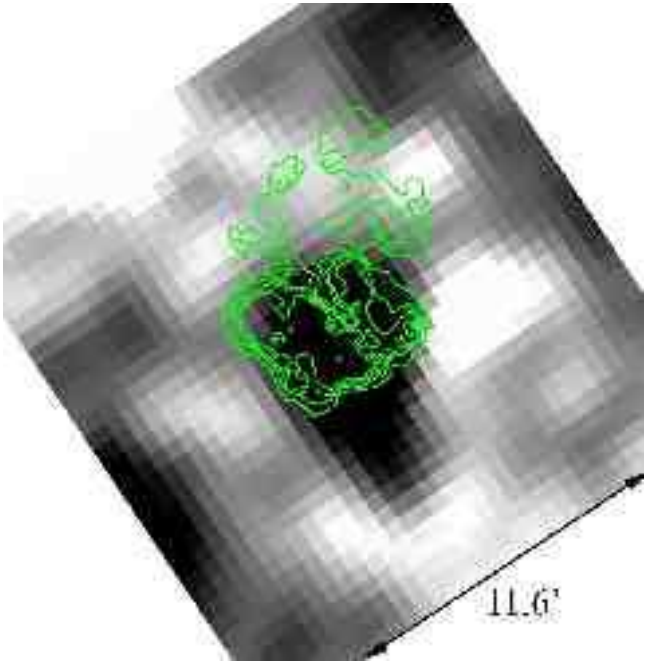}
  \caption{ H\,{\sc{i}} emission (in white) from the SGPS integrated between $-$15.66~km s$^{-1}$
  and $-$10.72~km s$^{-1}$.
  H$\alpha$ emission is superimposed as green contours}
  \label{hi}
\end{figure}

The \HI\ emission from the Southern Galactic Plane Survey (SGPS,
McClure-Griffiths et al. \cite{mcc05}) is shown in Fig.~\ref{hi},
integrated between $-$10.72~km s$^{-1}$ and $-$15.66~km s$^{-1}$.
The angular resolution of these observations is 40$\arcsec$ and
the velocity resolution is 0.82~km s$^{-1}$.
%fig11
The emission shows an annular structure. The \HII\ region and
condensations 1 and 2 lie in the direction of the central hole.
The annular structure is larger near $-$11~km s$^{-1}$ (the
systemic velocity) than near $-$15~km s$^{-1}$, so the whole
structure appears to be in expansion. However, we only see the
approaching side of the cloud, as no emission is observed around
$-$5~km s$^{-1}$, as would be expected from the receding side of
the cloud. We have no explanation for this fact.

\section{Discussion}
\subsection{Star-forming processes at work in RCW~120\label{sf}}
Several star-forming processes may be at work in the direction of
RCW~120.

\begin{figure}
\includegraphics[angle=0,width=90mm ]{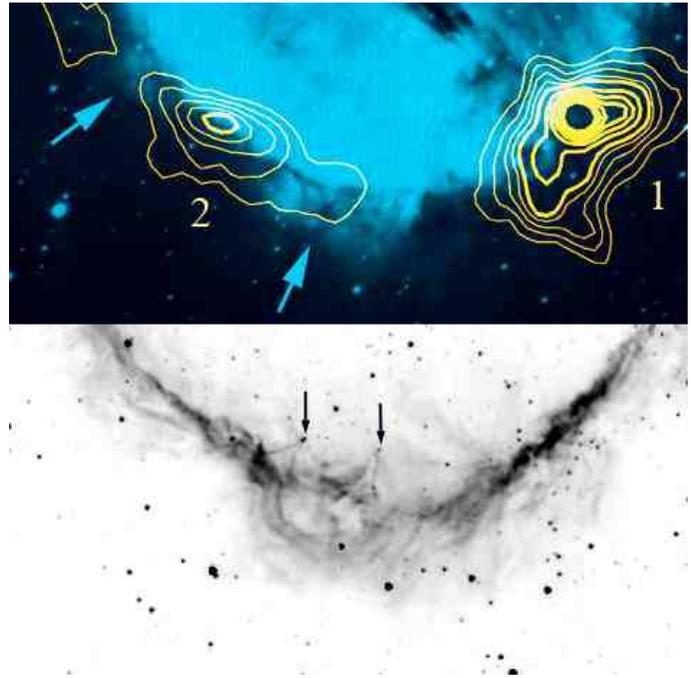}
  \caption{Instabilities of the ionization front. {\it Top}: the ionized gas traced by the H$\alpha$ emission (in blue) leaks between condensations 1 and 2,
  at the
  locations shown by the blue arrows.
  {\it Bottom}: structure of the ionization front (GLIMPSE 5.8\,$\mu$m image). The black arrows point to two YSOs
  (1\HII\ and 5\HII)
  observed at the vertices of triangular structures possibly created by dynamical instabilities in the ionization front }
  \label{inst}
\end{figure}

$\bullet$ The collect and collapse process: we observe a collected
layer of cold dust revealed by the 1.2-mm continuum emission. We
also observe fragments along this layer, suggesting that the
collected material has already experienced fragmentation. Fragments
1, 2, and 3 are clearly separated along this layer and the ionized
gas seems to be leaking between them (see Fig.~\ref{inst} top).
However, no massive star formation has been observed in the
direction of these fragments, contrary to what is observed on the
borders of Sh~104 and RCW~79. Note that the non-detection of an IR
source at the peak of the 1.2-mm emission could be due to the high
extinction we derive near this peak and to the low sensitivity of
the GLIMPSE observations. Deeper IR observations are needed to
possibly establish the presence of an IR star forming in this
condensation. The high-density peak observed in condensation~1 is
probably the best place to search for massive YSOs.

$\bullet$ The radiation-driven implosions of pre-existing clumps:
this is possibly the case for star formation associated with
condensation 4. The shape of the ionization front surrounding this
condensation is clearly distorted, protruding inside the \HII\
region. This situation is to be expected if the expanding
ionization front encounters a motionless dense clump. In this case
star formation may result from the radiation-driven implosion of
the clump (Lefloch \& Lazareff \cite{lef94}; Kessel-Deynet \&
Burkert \cite{kes03}).

$\bullet$ Dynamical instabilities of the ionization front: the
region located between condensations 1 and 2 is probably
dynamically unstable. Figure~\ref{inst} shows that the ionization
front is locally distorted, with the ionized gas leaking between
the condensations. We also observe two triangular structures
protruding into the ionized zone. Two YSOs, sources 1\HII\ and
5\HII, are observed at the vertices of these triangular structures
(see Fig.~\ref{inst} bottom). Such structures were possibly formed
by the development of instabilities in the ionization front, as
simulated by Garc{\'\i}a-Segura \& Franco (\cite{gar96}) and by
Dale et al.\ (in preparation). The latter simulation shows the
formation of stars at the vertices of such structures.

%fig12
\subsection{Star formation observed far from the ionization front \label{farsf}}
Figure~\ref{spacedist} presents the spatial distribution of the
Class I and Class II YSOs identified towards RCW~120. It shows that
many of these sources lie far from the ionization front -- as far as
1.5~pc. Some examples are the bright Class I sources 6out, 9out,
11out, 13out, 19out, 26out, 1C5, 5C6, and 12C1. The radial
structures clearly visible in absorption on the northeastern and
southwestern sides of the region (see Fig.~\ref{Figabs} right) are
striking evidence of interaction between the radiation leaking from
the IF and the surrounding medium.

In the absence of velocity information, it is impossible to be
sure that the observed YSOs are all associated with RCW~120. Only
a statistical approach can give us some insight into this point.
We have considered two zones located at the same Galactic latitude
as RCW~120 ($b=0\fdg499$) and on either side of it (at Galactic
longitudes $l=347\fdg16$ and $l=349\fdg16$). Each of these zones
has the same area as the region we searched for YSOs around
RCW~120 (Sect.~\ref{yso}). The same selection criteria was applied
to search for Class I and Class II sources in these zones as in
RCW~120 (i.e.\ $[3.6]-[4.5]$ and $[5.8]-[8.0]~\geq $0.4~mag). The
result is as follows: 15 and 25 YSOs, respectively, are detected
in these zones compared to 107 YSOs in the RCW~120 zone. This
suggests that most of the YSOs detected towards RCW~120 are
probably associated with it.

Which process can trigger star formation far from the ionization
front? Figure~\ref{Figabs} shows that low-brightness PAH emission
features also extend far from the ionization front. It seems that
the hot photon-dominated region, where PAH emission originates, is
more extended than would be expected if the ionization front were
impermeable. Examination of a wide-field H$\alpha$ image also
clearly shows extended H$\alpha$ emission surrounding the RCW~120
region. This suggests a leaky \HII\ region bounded by a porous
ionization front, allowing a fraction of the UV radiation to reach
regions far away from the front.

\begin{figure}
\includegraphics[angle=0,width=120mm ]{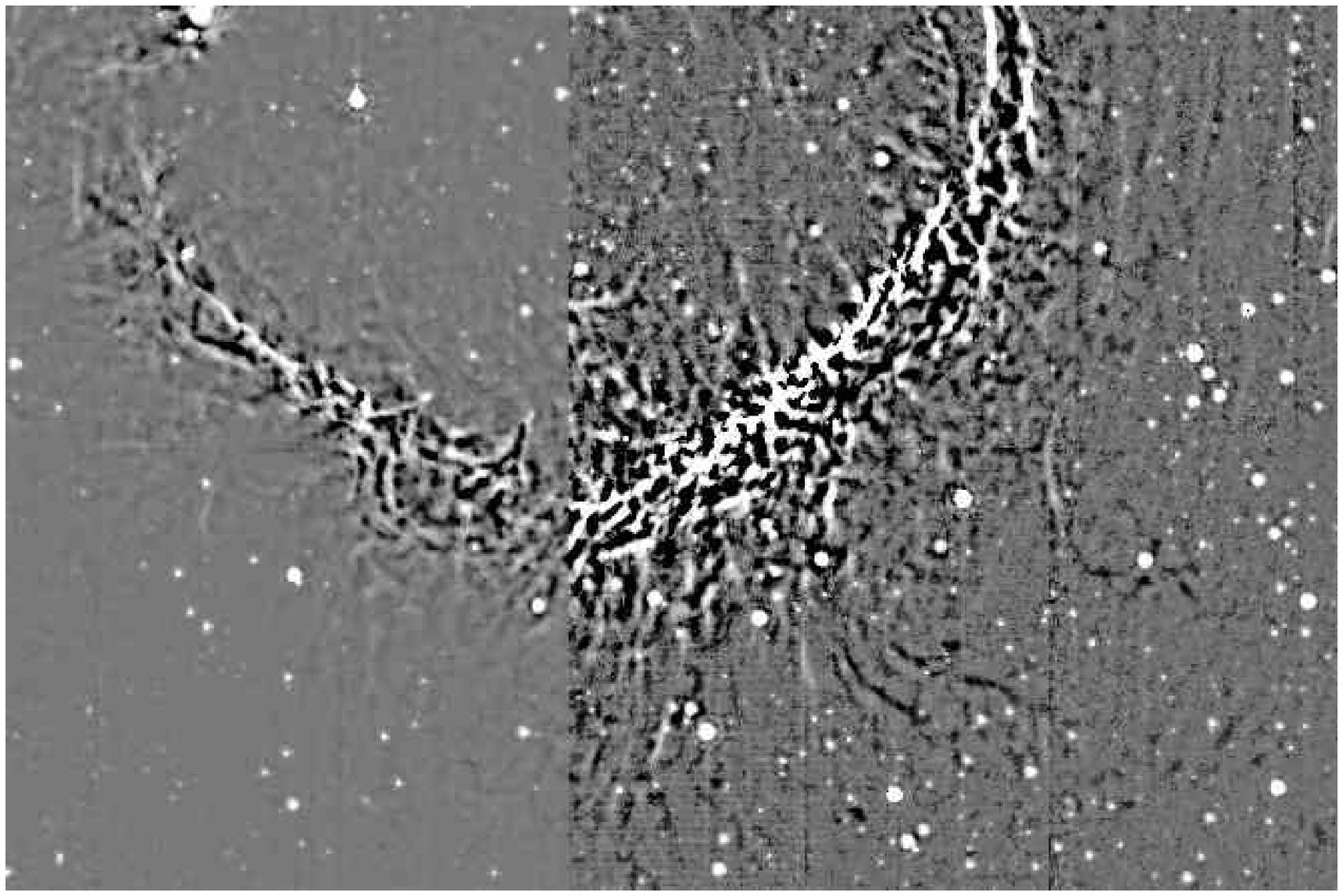}
  \caption{Unsharp-masked image of the southern part of RCW~120 at 8\,$\mu$m (see text for details)}
  \label{unsharp}
\end{figure}
%fig13
Figure~\ref{unsharp} is an unsharp-masked image of the southern part
of RCW~120 at 8\,$\mu$m, obtained by subtracting a median-filtered
version of the same image produced with a $12\arcsec \times
12\arcsec$ window from the original GLIMPSE image. The extended
bright structures are thus removed, leaving the low-brightness,
small-scale structures. The left part of Fig.~\ref{unsharp} shows
the highly irregular aspect of the IF. The right side, where the
contrast has been enhanced, shows the outer part of the PDR, with
many radial structures originating from the IF. This structure
clearly indicates that the UV radiation can penetrate far inside the
surrounding medium. Can the high pressure of the warm, partially
ionized inter-clump medium trigger star formation far from the
ionization front?

\subsection{Comparison with models}
The collect and collapse process of star formation, first proposed
by Elmegreen \& Lada (1977), has been formulated analytically by
Whitworth et al.\ (\cite{whi94}; see also Dale, Bonnell and
Whitworth \cite{dal07}) and simulated by Hosokawa \& Inutsuka
(\cite{hos05}, \cite{hos06}). All these models are spherically
symmetric around the exciting star and assume that the \HII\ region
expands into a uniform medium of density $n_0$. The assumption of
spherical symmetry is not too bad for RCW~120, but the observations
presented here indicate that the medium is far from homogeneous.
However, no model takes this fact into account.

A rough estimate of $n_0$ can be obtained by assuming that all the
material now observed, either ionized in the \HII\ region or neutral
in the massive fragments surrounding RCW~120, was previously located
in a sphere of density $n_0$ and of radius equal to that of the
\HII\ region. We used only the southern half of RCW~120 for this
estimate of $n_0$. From the radio-continuum flux density, we
estimate the mass of ionized gas to be $M$(\HII)~$=54~M_{\odot}$
(thus $27~M_{\odot}$ for half the \HII\ region). The mass of neutral
material in condensations 1, 2, 3, 4, and 7 is $633~M_{\odot}$ (the
maximum value, for $T$=20~K). The mass of neutral material in the
collected layer, but not detected in the direction of the centre of
the \HII\ region because its emission is below the sensitivity limit
of the 20~mJy/beam, is $\leq185~M_{\odot}$ (thus $\leq93~M_{\odot}$
for half the layer), hence an upper limit of $3000$~atoms~cm$^{-3}$
for $n_0$. A lower limit of $1400$~atoms~cm$^{-3}$ is obtained for
$T$=30~K.

The dynamical expansion of RCW~120 and its PDR can be analysed
using the model of Hosokawa \& Inutsuka (\cite{hos06}). A
one-dimensional, spherically symmetric numerical method is used.
The UV and far-UV radiation transfer, as well as the thermal and
chemical processes, are solved with a time-dependent hydrodynamic
code. We suppose a central star of $22~M_{\odot}$ (which,
according to Diaz-Miller et al.\ \cite{diaz98}, emits the same
number of ionizing photons as the exciting star of RCW~120) and a
uniform ambient density of 3000~atoms~cm$^{-3}$.

\begin{figure}
      \includegraphics[width=80mm]{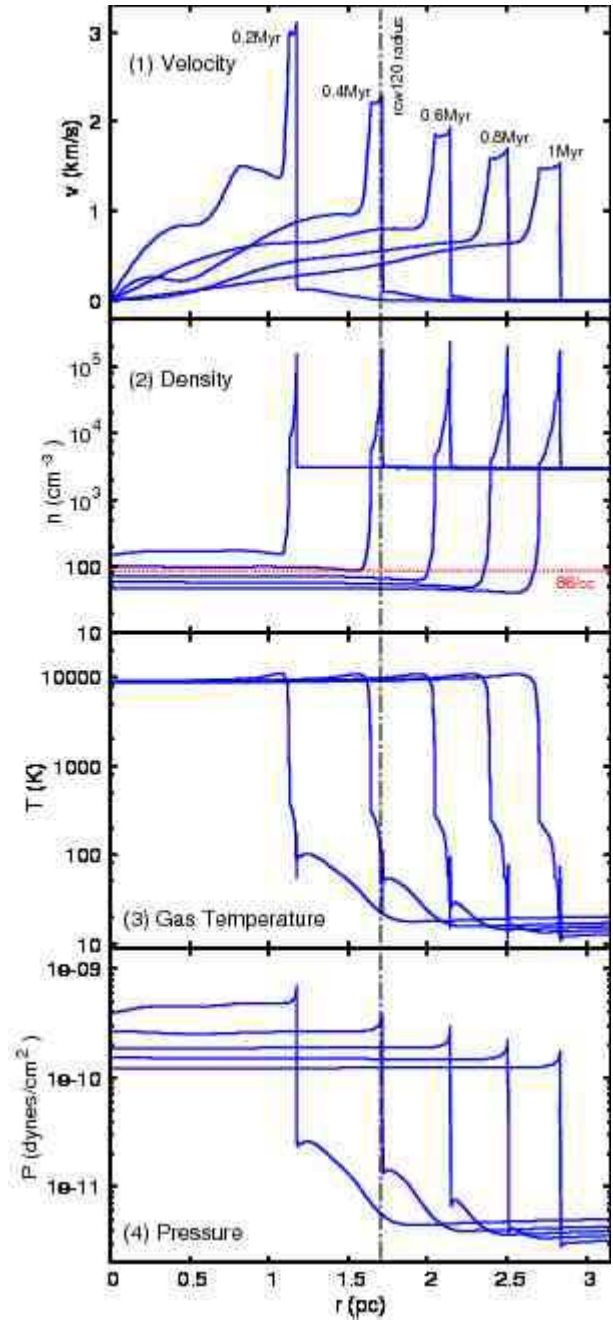}
      \caption{Snapshots of the gas dynamical evolution at $t$=0.2, 0.4, 0.6, 0.8, and 1.0~Myr}
      \label{fig1t}
\end{figure}

\begin{figure}
      \includegraphics[width=80mm]{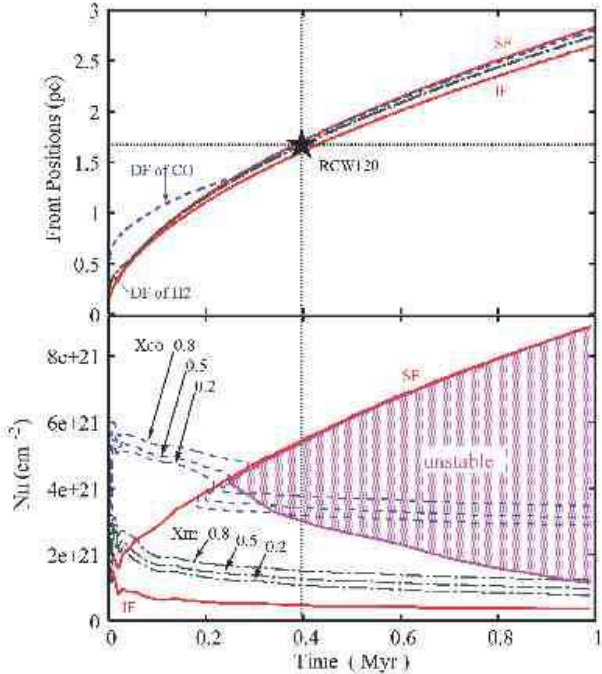}
      \caption{{\it Top:} Evolution of the positions of the various fronts (ionization front [IF], shock front
      [SF] and
      dissociation front [DF] of the H$_{2}$ and CO molecules). {\it Bottom:} Time evolution of the column density of the H$_{2}$ and
      CO  components.
      The shaded region corresponds to $t \geq (G \rho)^{-1/2}$ within the shell and hence
      indicates that gravitational fragmentation is expected. The contours show where the
      fractions of molecular gas $X_{\rm H_2}$ (dashed contours) and $X_{\rm CO}$ (dot-dashed contours) = 0.2, 0.5, and 0.8}
\label{fig2t}
\end{figure}

Figure~\ref{fig1t} shows the time evolution of several physical
quantities. The radius of RCW~120, 1.67~pc, is reached at
$t=0.4$~Myr. At this time the electron density of the ionized gas
agrees with the observed value of 86~cm$^{-3}$. Figure~\ref{fig1t}
also shows the  gas shell swept up by the shock front, with
densities in the range 10$^4$--10$^5$~cm$^{-3}$; the highest
density is found on the outside of the shell, close to the SF. The
shell mass is about $1000~M_{\odot}$ at this time, and most of the
swept-up hydrogen gas remains in the shell as hydrogen molecules.
%fig14
Figure~\ref{fig2t} shows the evolution of the positions of various
fronts. The swept-up shell very quickly becomes mainly molecular
(H$_2$); later on, at about 0.3~Myr, the dissociation front of the
CO molecule is engulfed by the expanding shell. Parts of the shell
become unstable when $t \geq (G \rho)^{-1/2}$. Figure~\ref{fig2t}
shows that the unstable region appears at about 0.3~Myr, near the
SF, and gradually spreads over the shell. Thus we would expect to
see, at the age of 0.4~Myr, a shell of mainly molecular collected
material, with the outside parts of the shell (near the SF)
fragmented. Of course, the model does not tell us if stars have
already formed in these fragments. Note that {\emph{these estimates
are rough}}, due to the uncertainties concerning both the density
and the uniformity of the medium into which RCW~120 evolves.
%fig15
\section{Conclusions}
Although it apparently has a very simple morphology, the \HII\
region RCW~120 elicits several questions.

It appears as an almost perfect sphere full of ionized gas -- a
Str\"omgren sphere around the exciting star. The sphere is open in
the direction of lower density, in the north, and the ionized gas
is escaping from the sphere. We are most probably seeing the very
beginning of a `champagne flow'. This is shown by the shape of the
\HII\ region, the shape of its IF as shown by the PAH emission and
by the velocity field.

Dense material, now mostly molecular, has been collected around the
ionized gas during the expansion of the \HII\ region. The collected
layer has begun to fragment. This is shown by the millimetre
emission of the cold dust. Some fragments are massive, but no
massive YSO is detected (up to 8\,$\mu$m at the GLIMPSE sensitivity)
in the direction of the fragments. If a massive Class~0 object is
forming, it may be detectable but only at longer wavelengths at the
emission peak of condensation~1.

Several Class~I and Class~II objects, of low and intermediate
mass, are observed in the direction of the PDR, near the IF. Their
formation was probably triggered by the expansion of the \HII\
region, via various processes such as dynamical instabilities of
the IF and the radiation-driven implosion of pre-existing
molecular clumps.

Two other points are pending and need further observations for
them to be understood:

-- What is the origin of the shell of atomic hydrogen surrounding
RCW~120? We suggest that it is part of the parental molecular cloud,
photo-dissociated from the outside by FUV background radiation
(Hosokawa \cite{hos07}). But why do we observe only half a shell of
\HI\ material (the half-shell approaching us)?

-- Several YSOs are observed far from the ionization front. Are they
associated with RCW~120? We have no proof that any specific YSO is
associated with RCW~120, but a statistical test suggests that most
of them are associated. If this is the case, what physical process
can trigger star formation far from the ionization front?

This point is very important and is probably linked to the structure
of the PDR. We are most probably dealing with a leaky \HII\ region,
in the sense that the ionization front is permeable to the UV
photons $(h\nu>$13.6~eV). This is shown by the presence of a
low-brightness H$\alpha$ zone surrounding RCW~120, which has the
same shape and almost the same extent as that of the low-brightness
PAH PDR. The low-density PDR, beyond the collected layer, appears to
be very irregular. If hydrogen-ionizing UV photons escape from the
\HII\ region, they ionize the low-density inter-clump medium,
destroying the PAH there. The photons may carve radial tunnels.
Lower energy photons heat PAHs at the surface of the high-density
neutral clumps. The pressure of this partially-ionized,
high-temperature, inter-clump medium may trigger star formation in
the low-mass clumps.

If real, such structures and such signatures of star formation
should be found around other \HII\ regions. Long-distance triggering
by radiation through a permeable medium should be investigated in
more detail by models and observations, as the results may change
our view of triggered versus spontaneous star formation.

\begin{acknowledgements}
      %------------------------
This research has made use of the Simbad astronomical database
operated at the CDS, Strasbourg, France, and of the interactive sky
atlas Aladin (Bonnarel et al.\ \cite{bon00}). Our long-term
collaborators on this project, B. Lefloch, J. Brand and F. Massi,
are warmly thanked for stimulating discussions. We thank R.~Cautain
for his help in creating the $K_{{\rm{S}}}$ mosaic image of RCW~120.
We thank the anonymous referee for important comments that helped to
clarify the text. This publication used data products from the
Midcourse Space EXperiment, from the Two Micron All Sky Survey, and
from the InfraRed Astronomical Satellite; for these we used the
NASA/IPAC Infrared Science Archive, which is operated by the Jet
Propulsion Laboratory, California Institute of Technology, under
contract with the National Aeronautics and Space Administration. We
also used the SuperCOSMOS H$\alpha$ survey. This work is based in
part on GLIMPSE data obtained with the Spitzer Space Telescope,
which is operated by the Jet Propulsion Laboratory, California
Institute of Technology, under NASA contract 1407.
\end{acknowledgements}

\end{document}